%% file: ms.tex
\documentclass[12pt,preprint]{aastex}

\slugcomment{DSF - 39/2003}

\shortauthors{Ilenia et al.} \shorttitle{Low Mass Primordial}

\begin{document}

\title{Evolution and nucleosynthesis of primordial low mass stars}

\author{Ilenia Picardi\altaffilmark{1,2,3}, Alessandro Chieffi\altaffilmark{4,5,6}, Marco
Limongi\altaffilmark{7,5,6}, Ofelia Pisanti\altaffilmark{1,2},
Gennaro Miele\altaffilmark{1,2}, Gianpiero
Mangano\altaffilmark{1,2}, Oscar Straniero\altaffilmark{2,8},
Gianluca Imbriani\altaffilmark{2,8}}

\affil{1. Dipartimento di Scienze Fisiche, Universit\'a di Napoli
"Federico II", Complesso Universitario di Monte S'Angelo, Via
Cintia, 80126, Naples, Italy}

\affil{2. Istituto Nazionale di Fisica Nucleare, Sez. Napoli,
Complesso Universitario di Monte S'Angelo, Via Cintia, 80126,
Naples, Italy}

\affil{3. Instituto de F$\acute{i}$sica Corpuscolar - C.S.I.C./ Universitat de
Val$\grave{e}$ncia Edificio Istitutos de Paterna, Apt 22085, E-46071, Valencia, Spain}

\affil{4. Istituto di Astrofisica Spaziale e Fisica Cosmica (CNR),
Via Fosso del Cavaliere, I-00133, Roma, Italy}

\affil{5. School of Mathematical Sciences, P.O. Box, 28M, Monash
University, Victoria 3800, Australia}

\affil{6. Centre for Astrophysics and Supercomputing, Swinburne
University of Technology, Mail Number 31, P.O. Box 218, Hawthorn,
Victoria 3122, Australia}

\affil{7. Istituto Nazionale di Astrofisica - Osservatorio
Astronomico di Roma, Via Frascati 33, I-00040, Monteporzio Catone,
Italy}

\affil{8. Istituto Nazionale di Astrofisica - Osservatorio
Astronomico di Collurania-Teramo, I-64100, Teramo, Italy}

\begin{abstract}

We discuss in detail the evolutionary properties of low mass stars
$\rm (M~\le~1~M_\odot)$ having metallicity lower than $\rm Z=10^{\rm -
 6}$ from the pre- main sequence up to (almost) the end of the
Asymptotic Giant Branch phase. We also discuss the possibility that
the large [C,N/Fe] observed on the surface of the most Iron poor star
presently known, HE0107-5240, may be attributed to the autopollution
induced by the penetration of the He convective shell into the H rich
mantle during the He core flash of a low mass, very low metallicity
star. On the basis of a quite detailed analysis, we conclude that the
autopollution scenario cannot be responsible for the observed chemical
composition of HE0107-5240.

\end{abstract}


\section{Introduction}

The idea that stars born shortly after the Big Bang are still {\it
alive} has such a profound impact on the human emotionality that their
evolutionary properties have been investigated in spite of the fact
that their formation in an environment deprived of metals (Z=0) is at
least not favored because of the lack of a strong coolant.

From a purely theoretical ground the fragmentation of primordial
clouds is still largely debated and unclear, so that the current
literature on the subject still ranges from papers in which
essentially massive stars form \citep{abn00} to papers in which a
significant production of low mass stars $\rm (M\sim1~M_\odot)$ is
possible \citep{nu02}. However the continuous discovery of
progressively more metal poor low mass stars constitutes a very good
excuse to trust the primordial formation of low mass stars and hence
to study their evolutionary properties.

The first computation of the Main Sequence (MS) properties of
primordial (Z=0) low mass stars has been performed, as far as we know,
by \cite{cp75} while the first evolution up to the He core flash has
been published by \cite{da82}. In the latter paper all the key
properties of the evolution of these stars were already clearly
outlined. In particular, the presence of a Red Giant Branch (RGB)
powered mainly by the PP chain rather than by the CNO cycle together
to its natural consequences: a dimmer core mass- luminosity relation
and a more external (in mass) ignition of the He burning (due to the
slower advancing of the H shell). In the same paper it was also
suggested the possibility of the ingestion of protons in the "active"
He (and C rich) convective shell and hence the possibility to produce
stars showing very large surface CN abundances. In a following paper
\cite{gd83} tried to perform similar computations but the formation of
a (fast growing) convective core towards the end of the central H
burning phase forced them to stop the computations. Such a strong
instability was not encountered in any further study. \cite{fih90}
fully reanalyzed the evolution of a primordial $\rm 1~M_\odot$ from
the MS up the He flash and described in detail the evolution of this
star in comparison with stars of the same mass but having a larger metallicity, 
i.e. $\rm Z=10^{- 4}$ and $\rm Z=2\times10^{-2}$. Since we will not
tempt to rediscover once again the "funny" properties of low mass zero
metallicity stars, we refer the reader interested in a deep knowledge
of the properties of these stars to that paper. One of the main
results found by \cite{fih90} was that the outer border of
the He convective shell, that forms at the He core flash, penetrates, unambiguously, 
the H-rich mantle. Such a result was discussed and interpreted as due to
the combined effects of the unusually external location of the He
ignition and to the low entropy barrier existing at the He-H interface
as well. The properties that follow the ingestion of the protons in
the He convective shell were firstly discussed by the same group in a
following paper \citep{hif90}. In particular they showed that the
convective shell splits in two (at the mass location where the local
burning timescale becomes comparable with the mixing timescale), the
He convective shell dumps out, the H convective shell eventually
merges with the convective envelope an hence that the envelope of the
star becomes strongly enriched in the products of both the H and He
burnings. In a third paper, \cite{fii00} recomputed several models of
low mass zero (and non zero) metallicity and mainly confirmed their
previous findings. By the way, tabulations of evolutionary tracks of
extremely metal poor stellar models have been published in those years
by \cite{cc93} and \cite{cct96}: the He core flash properties of zero
metallicity stars were not addressed at all in those papers.
\cite{wcss00} recomputed a few low mass zero metallicity models and
concluded very firmly that there isn't {\it any evidence} of mixing
between the He convective shell that develops at the He core flash and
the H rich envelope. In the following year the same group
\citep{scsw01} found the mixing between the He convective shell and
the H-rich mantle at the He core flash of a $\rm 1~M_\odot$
(Y=0.23,Z=0); they also studied the dependence of the occurrence of
the mixing on several parameter (equation of state, initial He
abundance, initial mass, diffusion efficiency, external pollution and
opacity). Their main result was that the occurrence of this mixing is
very robust with respect to all the changings but for the dependence on
the initial He abundance (an increase of $\rm Y_{\rm ini}$ from 0.23
to 0.24 inhibits the mixing) and on the possible pollution of the
envelope (that also inhibits the mixing). For the smaller masses ($\rm
0.8~M_\odot$) they always found the mixing to occur! In a further
paper \citep{sscw02}, they extended the tests to the dependence on the
metallicity and found that the mixing occurs up to a metallicity of $\rm
Z<10^{\rm -6}$ for masses of the order of $\rm 0.8~M_\odot$. However,
other similarly recent computations by \cite{mgcw01} and \cite{sll02}
did not find any mixing between the He convective shell and the H rich
mantle at the onset of the He core flash.

In the last years we have continuously worked on the evolution of zero
metallicity stars \citep{cdls01,cl02,lc02,blc03,lcb03} in a wide mass
range that goes from the low masses up to the more massive ones ($\rm
100~M_\odot$). We have discussed over the years the properties of both
the primordial AGB and core collapse supernovae. In this paper we
present our low mass models of extremely low metallicity and discuss
the fit to the most Iron poor star presently known, HE0107-5240
\citep{chris02}. The paper is organized as follows. In section 2 we
briefly remind the current setup of the evolutionary code FRANEC
(release 4.98); section 3 is devoted to the description of the
properties of our "reference" stellar model, i.e. a $\rm 0.8~M_\odot$
having Y=0.23 and Z=0. Additional test models are discussed in section
4 while the fit to HE0107-5240 is discussed in section 5.

\section{Stellar evolution code and input physics}

\begin{figure}
\figurenum{1} \epsscale{1.0} \plotone{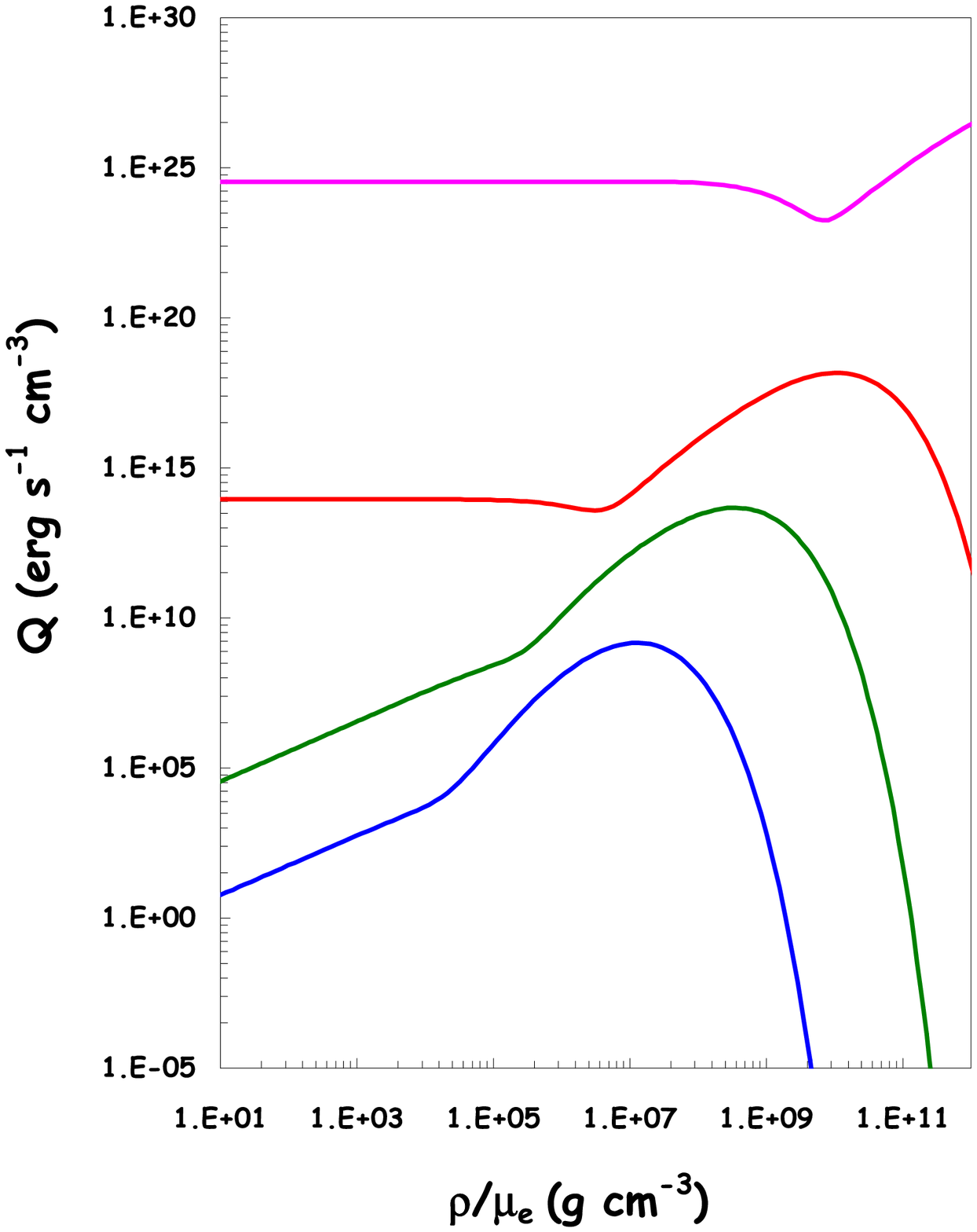}

\figcaption{The total energy loss rate in neutrino emission is
plotted versus $\rho/\mu_e$. From bottom to top the lines
correspond to $T=10^8$, $10^{8.5}$, $10^{9}$, $10^{10} \,K$
\cite{EMMPP02,EMMPP03}.\label{f00}}

\end{figure}

All the models presented in this paper have been computed by means of the FRANEC
(Frascati RAphson Newton Evolutionary Code) which is now at the release 4.98. The main
properties of this version of the code have been discussed in detail in \cite{cls98} and
will not be repeated here. One crucial updated has been included in the present setup of
the code, i.e. an update expression for the total energy loss rate, $Q$, due to neutrino
emission \citep{EMMPP02,EMMPP03}. This quantity is plotted in Figure \ref{f00} versus
$\rho/\mu_e$ for $T=10^8$, $10^{8.5}$, $10^{9}$, $10^{10} \,K$. The new evaluation of $Q$
takes into account the leading leptonic processes: pair annihilation,
$\nu$-photoproduction, plasmon decay, bremsstrahlung on nuclei up to the first order in
radiative corrections (see \citep{EMMPP02,EMMPP03} for details).

The network we have used in the present computations includes 44
nuclear species (from P to $\rm ^{\rm 30}Si$) interacting through
268 nuclear processes. The same network has been used for both the
H and He burning phases.

\section{The Evolution of the ${\rm \bf 0.8~M_\odot}$}

The typical mass we are interested in is the ${\rm 0.8~M_\odot}$
because it roughly corresponds to the mass born soon after the Big
Bang and evolving now off the Main Sequence. Its evolutionary
properties up to the end of the central He burning phase may be
divided in six parts: Main Sequence, first climbing along the RGB,
first He core flash, second RGB, second He core flash and central He
burning. Let us discuss the various phases in sequence. The full path
followed by this star in the HR diagram is shown in Figure \ref{f01}.

\subsection{From the MS to the He core flash}

\begin{figure}
\figurenum{3} \epsscale{1.0} \plotone{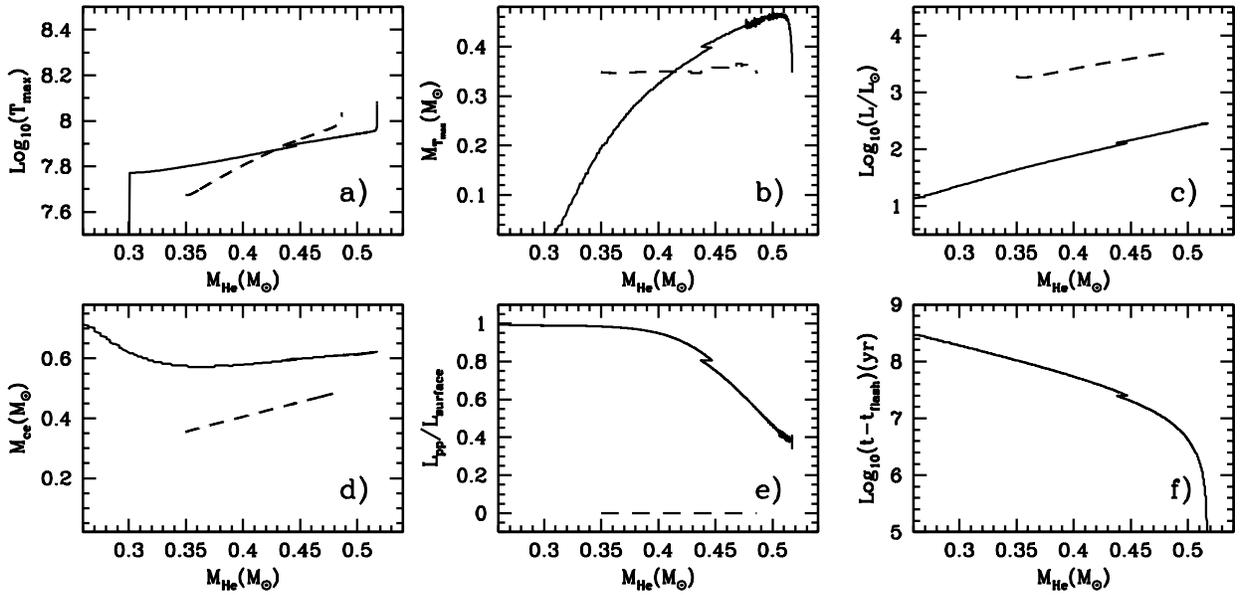}

\caption{Key properties of the evolution of a ${\rm 0.8~M_\odot}$ (Y=0.23 and Z=0)
along the RGB. The solid lines refer to the first climbing while the dashed ones
refer to the second climbing after the chemical composition of the envelope has
been polluted by the products of the He-H burnings.\label{f02}}

\end{figure}

\begin{figure}
\figurenum{4} \epsscale{1.0} \plotone{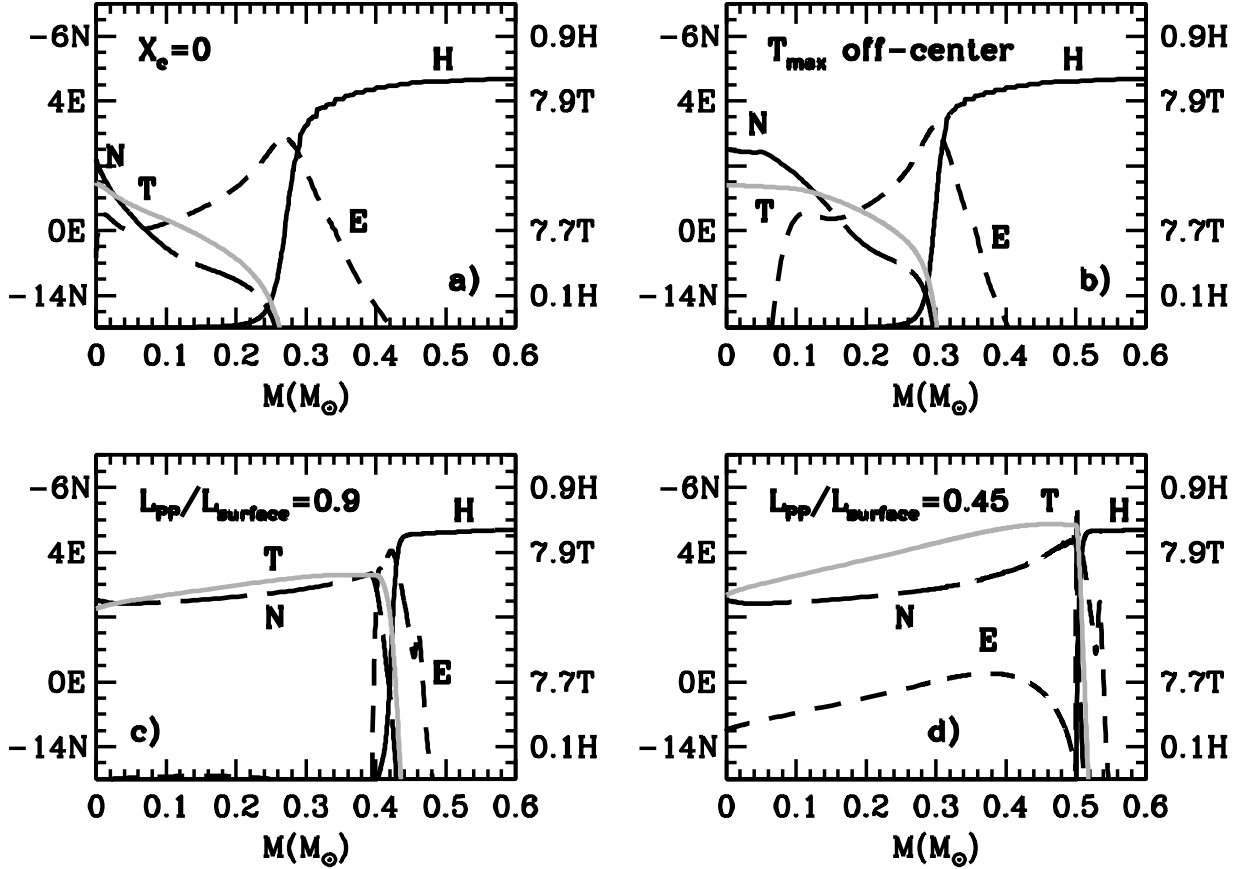}

\caption{Snapshot of the internal structure of a ${\rm 0.8~M_\odot}$ (Y=0.23 and
Z=0) at four characteristic points: a) the central H exhaustion, b) the off-center
shift of the maximum temperature, c) the model in which the global nuclear
luminosity produced by the PP chain drops to 90\% of the surface luminosity and d)
the model in which the global nuclear luminosity produced by the PP chain drops to
45\% of the surface luminosity. Each line has its properly labeled axis: H, T, N
and E stand, respectively, for Hydrogen (by mass fraction), Temperature (Log
scale), Nitrogen (by mass fraction and in Log scale) and nuclear Energy generation
rate (Log scale). \label{f03}}

\end{figure}

The key evolutionary properties of this star are shown in Figures
\ref{f02} and \ref{f03}. The various panels in Figure \ref{f02} show,
as solid lines, the runs of the maximum temperature (a), its mass
location (b), the surface luminosity (c), the mass of the convective
envelope (d), the fraction of luminosity produced by the PP chain (e)
and eventually the time spent above a given luminosity (f) as a
function of the He core mass ($\rm M_{\rm He}$). For sake of clearness
let us remind that ${\rm M_{\rm He}}$ has been defined as the mass
location corresponding to the maximum nuclear energy production by the
H burning. Figure \ref{f03} shows, in the four panels, snapshots of the
internal structure at various key points: the central H exhaustion
(a), the model in which the maximum temperature firstly shifts off
center (b) and the models where the ${\rm L_{\rm pp}}$ drops to 90\%
(c) and 45\% (d) of the surface luminosity. All the panels in Figure
\ref{f02} show the signature of the low efficiency of the PP chain
compared to that of the CNO cycle. The run of the maximum temperature with
the He core mass (panel a) shows a very high zero point (i.e. the
first maximum temperature at which it moves off center) because the PP
chain needs much higher temperatures than the CNO cycle to compensate
for the lower efficiency of the energy production and a very shallow
slope because of the low H burning rate. Such a low burning rate shows
up clearly in panel b) where the reduced accretion rate of fresh He on
the He core mass slows down the heating of the core (which is due to
the continuous growth of the He core mass) so that the energy losses
due to the neutrino emissions push (unusually) outward the mass
location of the maximum temperature (panel b). Accordingly, also the
luminosity (that is proportional to the accretion rate) is unusually
low (panel c). The overall higher temperature that is typical of the
metal free stars forces also the base of the convective envelope
(panel d) to remain much more external in mass than in "standard"
metal poor low mass stars. Panel e) shows that the PP chain provides
almost 100\% of the luminosity up to a He core mass of ${\rm
0.4~M_\odot}$ and that it progressively reduces down to roughly 40\%
at the onset of the He flash. This progressive reduction of ${\rm
L_{pp}}$ is due to the continuous increase of the CNO catalysts (and
associated ${\rm L_{CNO}}$) produced by a partial activation of the
${\rm 3\alpha}$ processes. By the way, the small hook that appears at
${\rm M_{\rm He}\simeq 0.44~M_\odot}$ marks the moment when the
maximum energy generation rate shifts from a layer dominated by the PP
chain to a (more internal) one where the CNO cycle prevails. The last
panel f) also reflects the slowness of the evolution along the RGB
when the nuclear energy production is dominated by the PP chain. The
four panels in Figure \ref{f03} show how the energy production and the
CNO catalysts change during the evolution along the RGB. Panel a)
shows the structure close to the moment of the central H exhaustion
$\rm (X_c=10^{\rm -6})$ and, in particular, the H profile (solid
line), the nuclear energy production  (short dashed line), the
Nitrogen abundance, which is largely the most abundant of the CNO
nuclei when it is at its full equilibrium, (long dashed line) and
eventually the temperature (gray line). Each quantity has its properly
labeled scale on either the primary or secondary Y axis. Though most
of the central H-burning phase is largely unaffected by the lack of
CNO catalysts because these low mass stars are anyway mainly powered
by the PP chain, the transition from the central to the shell H-
burning is deeply influenced by the lack of CNO nuclei. Since the CNO
cycle can not substitute the PP chain as main energy producer, the
core of a metal free star at the central H exhaustion is much hotter
than that of a "standard" metal poor star and at the central H
exhaustion the maximum nuclear energy production is already located at
roughly ${\rm 0.3~M_\odot}$ from the center. The central C mass
fraction produced by a partial activation of the $\rm 3\alpha$
processes (and fully converted in N) at the central H exhaustion is
${\rm 3\times10^{- 10}}$ dex. The other three panels show the same
quantities shown in panel a) at different times: it is evident how the
thickness of the H burning shell decreases with time and that the
global amount of CNO catalysts increases progressively up to ${\rm
3\times10^{- 8}}$ dex by mass fraction at the He core flash. Before
closing this section let us remind that the He flash starts at a mass
location significantly more internal than that of the maximum
temperature because the rate of the ${\rm 3\alpha}$ process depends
also on the square of the density. For this reason when the He core
flash starts, the maximum temperature quickly shifts where the rate of
the ${\rm 3\alpha}$ process has its maximum (see panel b in Figure
\ref{f02}).

\subsection{The "hot" encounter between He and H}

\begin{figure}
\figurenum{5} \epsscale{.6} \plotone{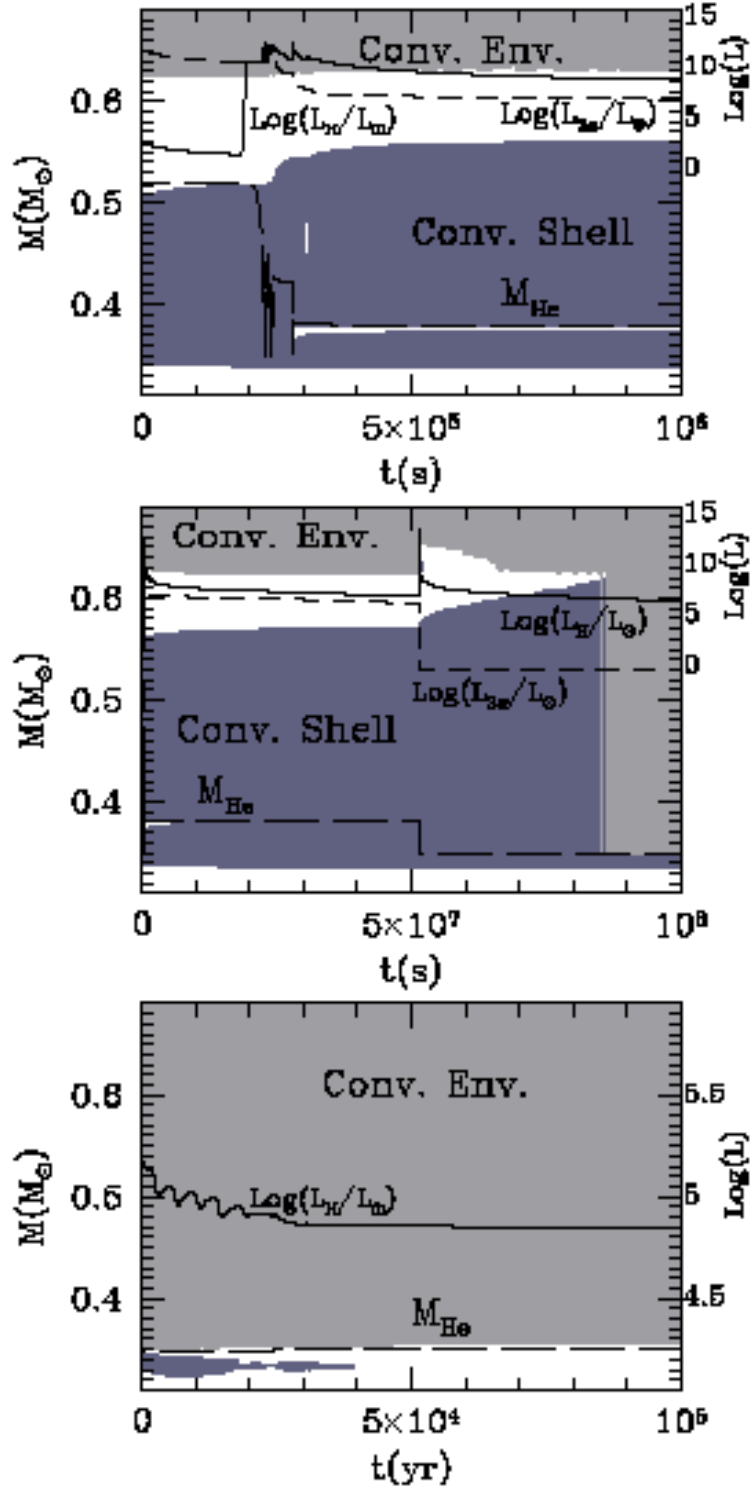}

\caption{Run of the convective regions as a function of time: the solid and short
dashed lines refer to the H and He luminosities, respectively (secondary axis)
while the long dashed line marks the location of the maximum H burning rate ($\rm
M_{\rm He}$). The light gray region marks the convective envelope while the darker
gray region(s) shows the convective shells. The temporal scale starts from the
maximum $3\alpha$ luminosity. The three panels show different temporal zooms: the
first $\rm 10^6$ s (upper panel), the first $\rm 10^8$ s (middle panel and the
first $\rm 10^5$ yr (lower panel). \label{f04}}

\end{figure}

When the $\rm 3\alpha$ luminosity ($\rm L_{\rm 3\alpha}$) raises above
$\rm 0.652~L_\odot$, a convective shell forms at a mass coordinate of
${\rm 0.348~M_\odot}$. 723 yr after the formation of this convective
shell, $\rm L_{\rm 3\alpha}$ reaches its maximum $\rm
(4.32\times10^{\rm 7} L_\odot)$ and the convective shell extends up to
$\rm M=0.506~M_\odot$. $\rm 2.1\times10^5$ s after $\rm L_{\rm
3\alpha}$ has reached its maximum, the still advancing He convective
shell reaches the tail of the H rich mantle and begins to dredge down
protons. At this stage the C mass fraction in the He convective shell
is $\rm 4.15\times10^{\rm - 2}$ dex while the total amount of C
produced by the He burning is $\rm 7.35\times10^{\rm - 3}~M_\odot$.
The following history of the mixing of protons in the He convective
shell is clearly described by the three panels in Figure \ref{f04}
where the solid and short dashed lines show the H and He luminosities,
the long dashed line marks the location of the maximum H burning rate
(the He core mass) while the light and dark gray zones show the
convective envelope and the convective shells, respectively. The three
panels show a zoom of the first $\rm 10^6$ s, $\rm 10^8$ s and $\rm
10^5$ yr after $\rm L_{\rm 3\alpha}$ has reached its maximum. The
protons ingested by the advancing He convective shell sink inward
without interacting with the surrounding environment until they reach
layers hot enough that their burning lifetime (against their capture
by $\rm ^{\rm 12}C$ nuclei) becomes shorter than the crossover time.
In our models such condition occurs at $\rm M=0.43~M_\odot$, where the
local temperature is $\rm\simeq115\times10^6$ K, and roughly $\rm
4\times10^4$ s after the beginning of the H ingestion. At this mass
location an H-burst producing a local luminosity maximum and a (very
mild) temperature inversion occurs: as a consequence the He convective
shell splits in two, a more internal one powered by the slowly dimming
$\rm L_{\rm 3\alpha}$ and a second one powered by the H burning. This
configuration lasts $\rm \simeq 2.8\times10^4$ s; since the electrons
are not strongly degenerate in the region where the H-burst occurs
($\rm \psi\simeq-2$), a rather quickly expansion quenches the energy
production so that the two convective shells merge and the protons
start flowing inward again. When they reach $\rm M=0.38~M_\odot$, the
local temperature is $\rm\simeq124\times10^6$ and a new H-burst,
followed by a splitting of the convective shell, occurs. This second
splitting phase lasts $\rm 5\times10^{\rm 7}$ s until the expansion
and cooling induced by the H-burst itself allows once again the
merging of the two convective shells. The further penetration of the
H-rich matter is not halted any more down to the base of the He
convective shell ($M\simeq0.35~M_\odot$). The He burning is now
completely switched off and the star readjusts quickly on a typical
"standard" RGB configuration with one shell energy source. Roughly 1
yr after the protons have reached the region where the He flash
started, the H-rich convective shell and the convective envelope merge
and the product of the He-H burnings are fully mixed homogeneously up
to the surface of the star (table 1, column 1, rows 1 to 17). The
nuclei (other than He) that are brought to the surface are mainly C, N
and O because the main result of the ingestion of protons is that of
redistributing the $\rm ^{\rm 12}C$ produced by the He burning among
the CNO nuclei via the CNO cycle. The global metallicity (i.e. the sum
of all the nuclei with $A>4$) is now $\rm 1.8\times10^{\rm -2}$ dex.
By integrating over all the H-rich layers one obtains a global
abundance of elements heavier than He equal to $\rm 8.08\times10^{\rm
- 3}~M_\odot$. It is worth noting that this value corresponds almost
exactly to the amount of C produced by the He shell up to the moment
when the He convective shell starts penetrating the H-rich mantle (as
number of particles, obviously). In other words the total number of
particles produced by the He flash up to the beginning of the H
ingestion remains freezed up to the onset of the 2nd He core flash
(see below). Another isotope that is largely produced is $\rm ^{\rm
7}Li$ that reaches a surface abundance of $Log(\epsilon)\simeq5.87$,
i.e. this star appears as an iper Li rich star. Such an enormous
production of Li occurs in two successive steps: during the three H-
 bursts a large amount of $\rm ^{\rm 7}Be$ is produced and distributed
within the convective shell. Then, the expansion and cooling induced
by the H-burst favors the $\rm ^{\rm 7}Be(e^- ,\nu)^{\rm 7}Li$
production, and the final merging of the convective shell with the
convective envelope redistribute the $\rm ^{\rm 7}Li$ within the
mantle of the star up to the surface. Summarizing, the ingestion of H-
 rich matter in the He convective shell deeply influences the further
evolution of the star: the H-bursts triggered by the activation of the
CNO cycle at very high temperatures induce an expansion of all the
region down to the ignition point of the $\rm 3\alpha$ process and
hence the He core flash fails. Moreover, the C produced by the He
burning is reconverted in C, N and O via the CNO cycle and the $\rm
^{\rm 13}C(\alpha,n)^{\rm 16}O$. $\rm ^{\rm 7}Li$ is also produced in
huge amount. The failure of the He core flash summed to a large
reduction of the He core mass pushes back the star towards a {\it deja
vu} configuration in which an active H-burning shell located at the
surface of the He core mass sustains the H rich mantle, i.e. an RGB
configuration. The transition towards this kind of structure
leads to the formation of a typical convective envelope extending
from the surface of the star down to close the active H burning shell
and hence triggers the merging of the H convective shell with the
outer convective envelope.

\subsection{The 2nd RGB and the 2nd He core flash}

\begin{figure}
\figurenum{2} \epsscale{1.0} \plotone{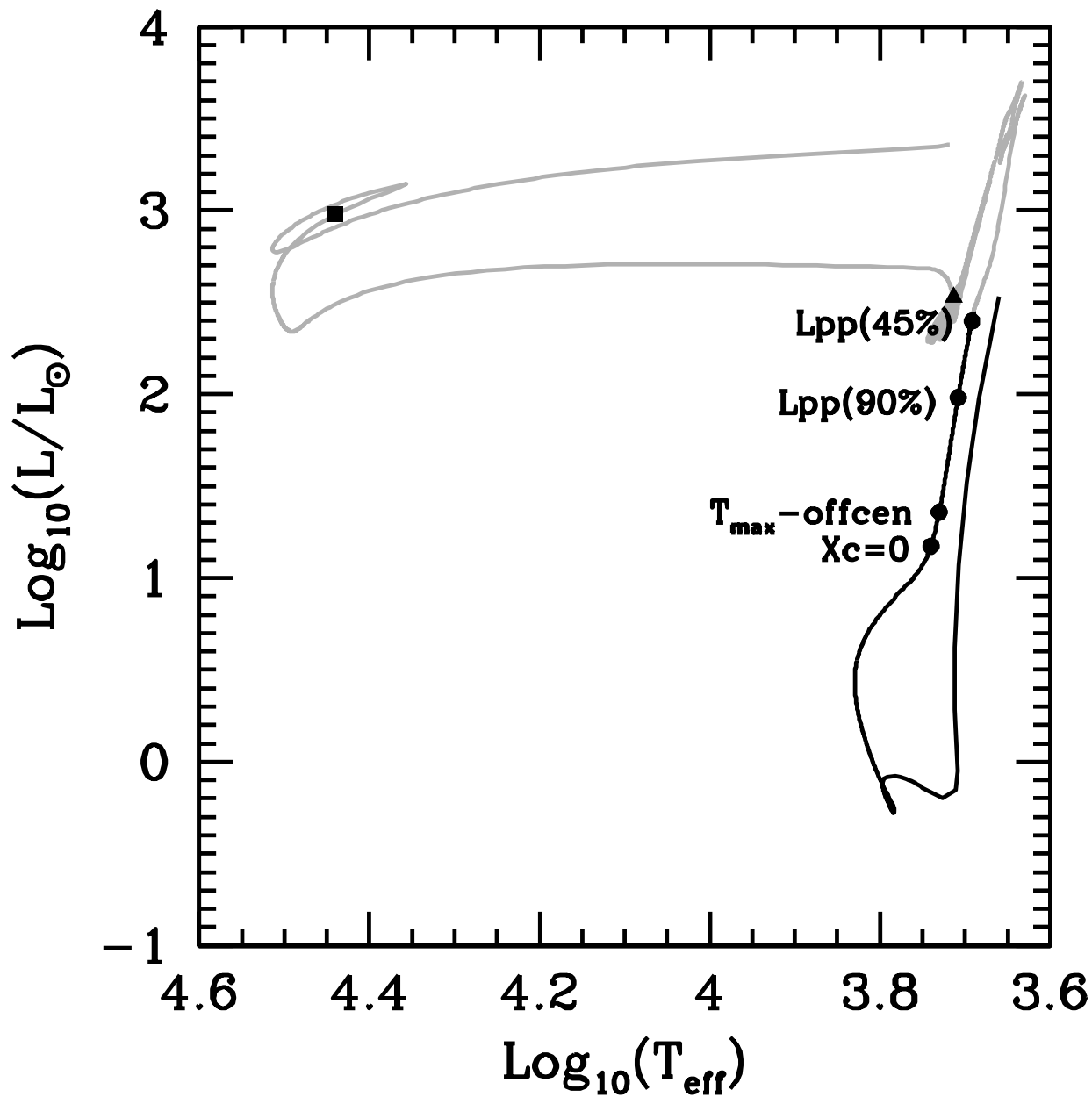}

\figcaption{Evolution of a ${\rm 0.8~M_\odot}$ (Y=0.23 and Z=0) from the
contraction phase along the Hayashi track up to the end of the central He burning.
The black line refers to evolution up to the first core He flash while the gray
line corresponds to the further evolutionary phases when the surface chemical
composition is polluted with the elements dredged to the surface. A few key points
along the RGB are marked by filled dots, the beginning of the central He burning is
marked by a filled triangle while the central He exhaustion is marked by a filled
square.\label{f01}}

\end{figure}

Roughly $\rm 2\times10^5$ yr after the beginning of the first He core
flash, the star completes its readjustment, settles once again on the
RGB and starts climbing it again towards a second He core flash. Its
structure is characterized by a helium core mass of
$\rm M_{\rm He}=0.357~M_\odot$ and by a
convective envelope whose chemical composition is largely enriched in
He (see table 1, column 1), CNO $\rm (Z_{\rm CNO}=1.8\times10^{\rm -
 2})$ and Li. It is worth recalling at this point that in the Sun $\rm
Z_{\rm CNO}=1.44\times10^{\rm -2}$, i.e. the initially metal free star
has an envelope super metal rich in terms of $\rm Z_{\rm CNO}$. Such
an occurrence obviously has a deep consequence on the behavior of the
star along the second RGB. The dashed lines in Figure \ref{f02} refer
to this phase. Since the H- burning shell now advances very quickly
because of the combined effects of the very high $\rm Z_{\rm CNO}$ and
of the low H mass fraction, it follows that: the maximum temperature
increases now very steeply with the He core mass so that the size of
the He core mass at the He flash is just $\rm 0.486~M_\odot$ (panel
a), the mass location of the maximum temperature remains at $\rm
M=0.36~M_\odot$ because of the very fast advancing of the H-burning
shell (panel b), the surface luminosity (panel c) reflects the much
larger H- burning rate, the base of the convective envelope closely
approaches the H shell (panel d) and the $\rm L_{\rm PP}$ contributes
negligibly to the total energy production (panel e). It follows that
the time spent by the star on this 2nd RGB phase is extremely short,
of the order of 2 Myr. The second RGB extends up to $\rm
Log(L/L_\odot)=3.726$ (see Figure \ref{f01}).

\begin{figure}
\figurenum{6} \epsscale{1.} \plotone{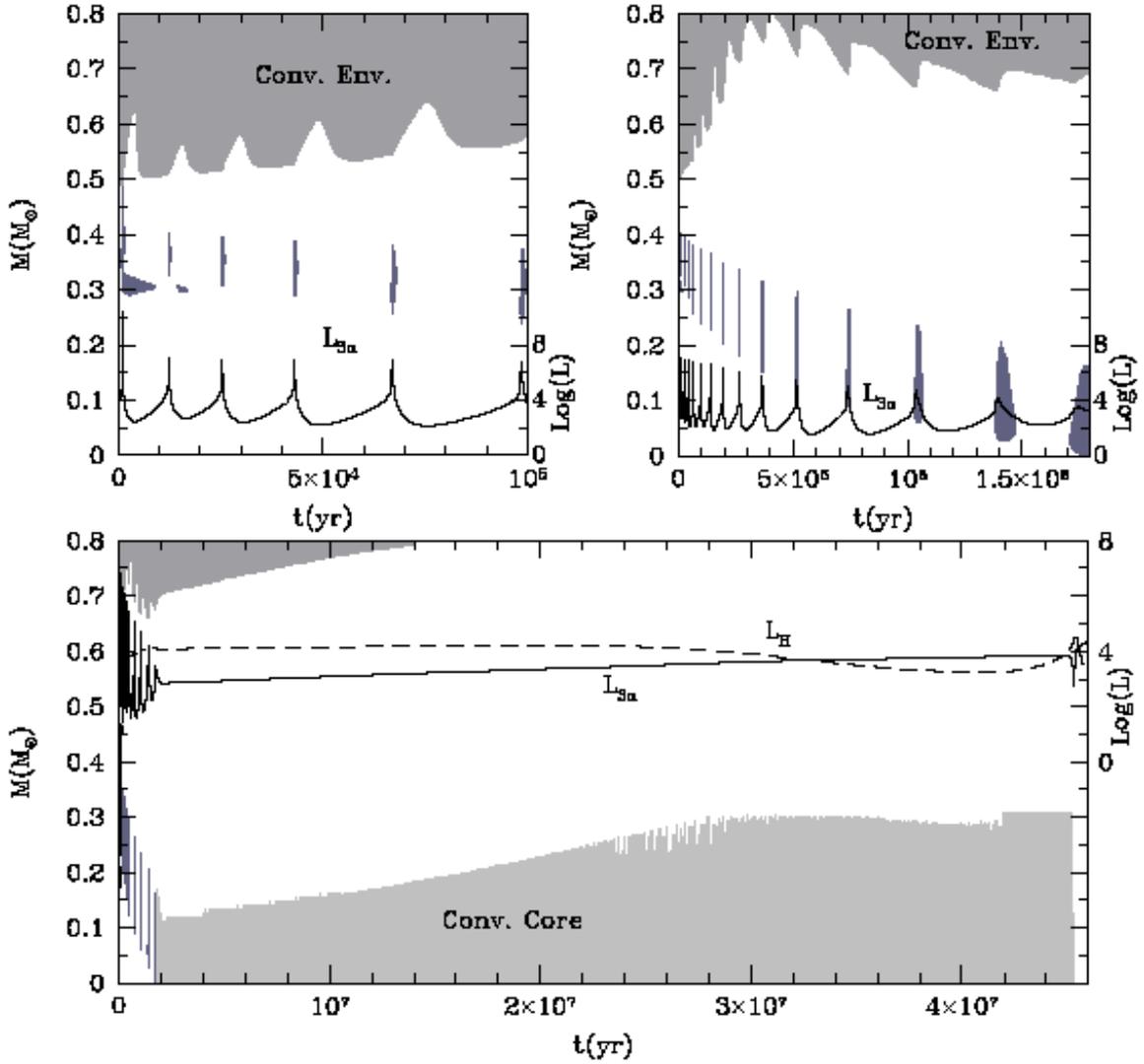}

\caption{Second flash and central He burning. The temporal scale is counted since
the formation of the He convective shell. The solid lines refer to the $\rm
3\alpha$ luminosity while the dashed line refers to the H luminosity. The dark gray
region marks the convective envelope while the light gray zones show either the
convective shells and the convective core during the central He burning.
\label{f05}}

\end{figure}

When $\rm M_{\rm He}$ reaches $0.486~M_\odot$, the maximum temperature
is high enough to trigger once again the $\rm 3\alpha$ process and a
second He core flash starts. Though the ignition point is very similar
to the one where the first He core flash occurred, this time the
entropy barrier at the H-He interface is large enough to prevent the
penetration of the outer border of the He convective shell in the H
rich mantle. Also the energy released is much lower than in the first
He core flash, the $\rm L_{\rm 3\alpha}$ luminosity peak reaching $\rm
1.4\times 10^5 L_\odot$, i.e. roughly 400 times weaker than in the
previous flash. The He core flash develops as usual through a series
of successive He burning ignitions (and associated convective shell
episodes) that lift the electron degeneracy in regions progressively
closer to the center. This star experiences a series of 15 flashes
that are shown in Figure \ref{f05}. During this phase the star rolls
in the HR diagram towards lower luminosities and slightly larger
effective temperatures. Once the electron degeneracy is fully removed
in the He core, the star settles in the point marked by the filled
triangle in Figure \ref{f01} and $\rm 1.8\times 10^6 yr$ are passed
from the beginning of the second He core flash.

\subsection{The central He burning}

As shown long time ago \citep{sg76}, the presence, or absence, of an
extended blue loop on the Horizontal Branch (HB) phase is largely due
to the amount of He in the envelope: for Y=0.4 dex a very extended
blue loop occurs even for stars that start their HB evolution very
close to their Hayashi track. Our stellar model has Y greater than 0.5
dex and this easily accounts for the extended blue loop visible in
Figure \ref{f01}. As for the percentage time spent by the star at the
various effective temperatures, one third of the total central He
burning lifetime is spent at the red and one third at the blue side of
the HR diagram, the remaining third being spent crossing (on a nuclear
timescale) the HR diagram from the red to blue. Another thing worth
mentioning is that the C mass fraction in the He convective core at
the beginning of the central He burning is 0.06 dex. The HB luminosity
is apparently very high ($\rm Log(L/L_\odot)\simeq2.7$) and the HB
lifetime is very short ($\rm 4.5\times 10^7 yr$) for a star having an He
core mass of $\rm 0.486~M_\odot$. The reason for both these
occurrences is, once again, the very high $\rm Z_{\rm CNO}$ and the
low H abundance in the envelope that force the H burning shell to be
very efficient in piling up fresh He on the He core mass. By the time
the star reaches $\rm Log(T_{eff})=3.74$ (coming from the red) its He
core mass has already grown up to $\rm 0.6~M_\odot$ and then it
continues to increase up to $\rm 0.749~M_\odot$ at the central He
exhaustion (marked by a filled square in Figure \ref{f01}). All the
key properties of the evolution of this star are reported in the first
column of Table 1. After the mass, initial metallicity and He
abundance, rows 4 to 17 refer to the surface abundances of the most
abundant nuclei. The following rows show: the time spent on the first
RGB above $\rm Log(L/L_\odot)=2$; the He core mass, the mass location
of the maximum temperature and the surface luminosity at the He core
flash; the duration of the first He core flash; the He core mass, the
mass location of the maximum temperature and the surface luminosity at
the beginning of the second climbing on the RGB; the lifetime of the
2nd RGB phase; the He core mass, the mass location of the maximum
temperature and the surface luminosity at the 2nd He core flash; the
He core mass at the beginning and the end of the central He burning
phase; and eventually the central He burning lifetime.

\section{Additional models}

We have shown in the previous sections that the He core flash of a
$\rm 0.8~M_\odot$ of primordial chemical composition (Y=0.23 and Z=0)
is characterized by the penetration of the He convective shell in the
H rich mantle. Such a phenomenon has very interesting consequences,
the most important one being the dredge up of large amounts of C, N, O
and Li to the surface. If this peculiar phenomenon  were confined to
just this mass and to this specific chemical composition, its interest
would be largely academical. If, vice versa, it would occur for a
reasonably wide range of values of the three parameters (M, Y and Z),
its {\it reality} would be strengthened and the probability of its
occurrence in real stars would increase significantly. Hence, in order
to assess the robustness of this phenomenon, we have computed a few
further models, i.e. three models in which we have progressively
increased the initial metallicity ($\rm Z=10^{\rm -8},~10^{\rm -
 7}~and~10^{\rm -6}$), a further one computed with a higher initial He
abundance (Y=0.27) and other two models with a higher mass (1 and $\rm
1.5~M_\odot$).

As for the metallicity, we found that up to Z=$\rm 10^{\rm -7}$ the
evolutionary properties of the $\rm 0.8~M_\odot$ closely follow those
found for the metal free case (columns 2 and 3 in Table 1), so that
also for these metallicities the He convective shell penetrates the H
rich mantle. The final surface chemical composition after the dredge
up is quite similar, the global $\rm Z_{\rm CNO}$ mildly increasing
from $\rm 1.8\times10^{\rm -2}$ (Z=0) to $\rm 2.6\times10^{\rm - 2}$
($\rm Z=10^{\rm - 7}$). In all three cases the $\rm ^{\rm 12}C/^{\rm
13}C$ is very low (3-4) while the C/N ratio increases from 1.3 (Z=0)
to 5.3 ($\rm Z=10^{\rm -7}$) mainly because of the increase of the
Carbon abundance. The Oxygen abundance remains quite constant over
this metallicity range and the surface $\rm ^7Li$ abundance remains
very high up to $\rm Z=10^{\rm -7}$. The run with $\rm Z=10^{\rm -6}$
did not show any mixing at the He core flash between the He convective
shell and the H rich envelope and hence we followed this evolution
just up to the end of the first major He flash.

\begin{figure}
\figurenum{7} \epsscale{1.} \plotone{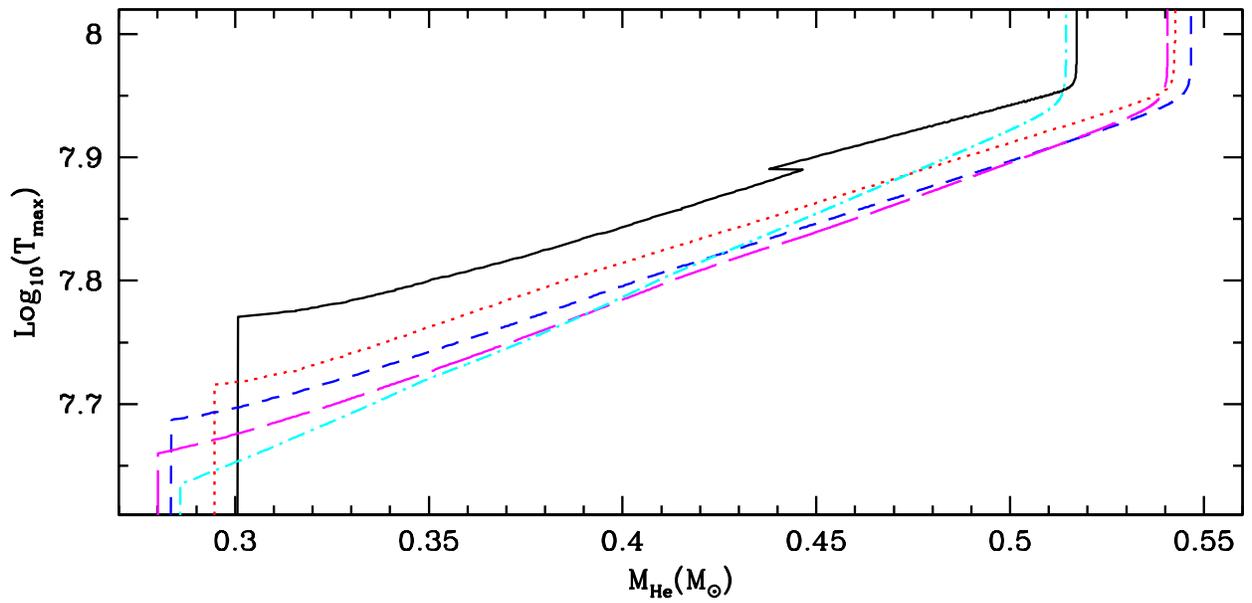}

\caption{Run of the maximum off center temperature versus the He core mass. The
various lines refer to 0.8 $\rm M_\odot$ (Y=0.23) stellar models having different
initial metallicities:Z=0 ({\it solid}), $\rm Z=10^{\rm -8}$ ({\it dotted}), $\rm
Z=10^{\rm -7}$ ({\it short dashed}), $\rm Z=10^{\rm -6}$ ({\it long dashed}) and
$\rm Z=10^{\rm -4}$ ({\it dot-dashed}). \label{f06}}

\end{figure}

A feature of these extremely metal poor stellar models is that the He
core mass at the He flash increases steadily with the metallicity up
to $\rm Z=10^{\rm - 7}$ and only above this value it start decreasing
as the metallicity increases (which is the well known trend for the
low mass stars at metallicities larger than $\rm Z=10^{\rm -4}$). The
explanation for such an occurrence can be easily understood by looking
at Figure \ref{f06} that shows the trend of the maximum (off-center)
temperature with the He core mass. The size of the He core mass at the
He flash is determined (in general) by both the slope and the zero
point of the $\rm Log(T_{\rm max})-M_{\rm He}$ (almost linear)
relation. The zero point (which in practice may be identified as the
temperature at which it starts moving off center) scales inversely
with the metallicity (actually $\rm Z_{\rm CNO}$) because the larger
the $\rm Z_{\rm CNO}$ the earlier the CNO cycle starts dominating the
nuclear energy production and hence the cooler will be the interior of
the star while it exits the Main Sequence to enter the RGB phase. The
slope of that relation, on the contrary, steepens increasing the
metallicity ($\rm Z_{\rm CNO}$) because the larger the metallicity the
faster the H burning shell advances and hence the faster will be the
growth of the He core and its heating. Figure \ref{f06} shows clearly
that an increase of the metallicity from 0 (solid line) to $\rm
Z=10^{\rm -8}$ (dotted line) and to $\rm Z=10^{\rm -7}$ (short dashed
line) lowers progressively the temperature at which it starts moving
off-center (the zero point) while it does not alter at all the slopes
of the three run. In other words the H burning shell is dominated by
the PP chain up to a metallicity of the order of $\rm Z=10^{\rm -7}$.
This means that between Z=0 and $\rm Z=10^{\rm -7}$ the final He core
mass is mainly determined by the "hotness" of the star when it leaves
the Main Sequence: the cooler the initial off center temperature the
larger the amount of mass that must be accreted in order to ignite the
$\rm 3\alpha$. As the metallicity increases to higher metallicities,
$\rm Z=10^{\rm -6}$ (long dashed line) and $\rm Z=10^{\rm -4}$ (dot
dashed line), the H burning shell begins to be dominated by the CNO
cycle that speeds up the RGB evolution (steepens the slope) and leads,
in turn, to smaller He core masses.

The influence of the initial He abundance on the He-H mixing at the He
core flash has been studied by computing a $\rm 0.8~M_\odot$ having
Z=0 and Y=0.27. Also in this case we found an extended mixing followed
by the dredge up to the surface of the product of the He-H burnings.
The surface chemical composition after the dredge up is shown in the
4th column of Table 1. The influence of the initial mass has been
studied by computing two stellar models of masses $\rm 1.0~M_\odot$
and $\rm 1.5~M_\odot$ (Y=0.23 and Z=0). Also the $\rm 1.0~M_\odot$
experiences a strong mixing between the He convective shell and the H
rich mantle (the resulting surface chemical composition is shown in
the 5th column of Table 1) while the $\rm 1.5~M_\odot$ ignites the He
quiescently without any He core flash. To stress simultaneously the
effect of the mass and the initial He abundance we have also computed
the evolution of a $\rm 0.9~M_\odot$ with Y=0.27 and Z=0. Also in this
case we found an extended penetration of the He convective shell in
the H rich mantle (this evolution was stopped as soon as the He
convective shell was definitely penetrated in the H rich envelope).
All these tests show that a) the He- H mixing at the He core flash is
a phenomenon that occurs over a reasonably large range of
metallicities, initial He abundances and masses and b) within the
range of parameters addressed in this paper, the final surface $\rm
Z_{\rm CNO}$ is only mildly dependent on the
initial mass, metallicity and He abundance.

Before closing this section let us briefly comment on the comparison
between our results and others available in the literature. In
particular we want to stress here that, in spite of the quantitative
differences that come out when closely comparing the models computed
by different groups, a few key features seem to be quite well
established (fortunately). First of all the sequence of events that
leads to the surface pollution: firstly, the penetration of the outer
border of the He convective shell in the H rich mantle, then, the
splitting of the He convective shell in two at the mass location where
the H-burst occurs and, eventually, the merging of the H-convective
shell with the convective envelope. What it is still more interesting
is that the amount of C and N brought to the surface is always of the
same order of magnitude (a few times $\rm 10^{\rm -3}$ dex) as well as
the C/N ratio that it is always roughly of the order of one (with a
maximum excursion of a factor of two). Also the limiting metallicity Z
(actually one should mention $\rm Z_{\rm CNO}$) that allows the He
convective shell to penetrate the H rich mantle is quite similar to
the one quoted by \cite{sscw02}, i.e. $\rm Z=10^{\rm -6}$ (that
corresponds to a scaled solar $\rm Z_{\rm CNO}=7.2\times10^{\rm -7}$).

\section{The HE0107-5240 case}

A large C and N overabundances relative to Iron is a quite common
occurrence among the extremely iron poor stars, e.g. \cite{b99}, and
hence it is quite tantalizing to associate such overabundances with
the "hot" He-H mixing that occurs exactly (and only) at the extremely
low Iron abundances. Two stars have been analyzed in detail up to now
\citep{scsw01,sscw02}: the first one, CS22892-052, has $\rm
[Fe/H]\simeq- 3$, $\rm [C/Fe]\simeq1.1$, $\rm [N/Fe]\simeq1$, and $\rm
^{\rm 12}C/^{\rm 13}C\ge10$, while the second one, CS22957-027, has
$\rm [Fe/H]\simeq- 3.4$, $\rm [C/Fe]\simeq2.2$, $\rm [N/Fe]\simeq2$,
and $\rm ^{\rm 12}C/^{\rm 13}C\simeq10$. Though these two stars are
indeed extremely Fe poor, their Fe abundance is large enough that even
a scaled solar $\rm Z_{\rm CNO}$ would bring them outside the range of
values for which the hot He-H mixing occurs (in the present generation
of models). A very possible O overabundance would even worsen the
situation. To overcome such a difficulty, \cite{scsw01,sscw02}
postulated that these stars were born by pristine material and that
they accreted enriched material only later on, during their Main
Sequence lifetime. However, even with this {\it escamotage}, the
analysis showed that there is a total inconsistency between the
observed and predicted surface abundances: in particular the predicted
[C,N/Fe] are exceedingly larger than the observed values.

Quite recently it has been discovered (and quite accurately analyzed)
the most iron poor star presently known: HE0107-5240 \citep{chris02}.
Its iron abundance is 1/200000 of that in the Sun ([Fe/H]=-5.3), C, N
and Na are enormously enhanced (by a factor of $10^4$, $10^{2.3}$ and
10, respectively) relative to iron, Mg is almost scaled solar, $\rm
^{\rm 12}C/^{\rm 13}C>30$ and $\rm [Li/Fe]<5.3$. Its extremely low
iron abundance and its high [C/Fe] and [N/Fe] make this star the ideal
candidate to check if these large surface overabundances may be due or
not to the "hot" mixing between He and H during the He core flash.

\begin{figure}
\figurenum{8} \epsscale{1.} \plotone{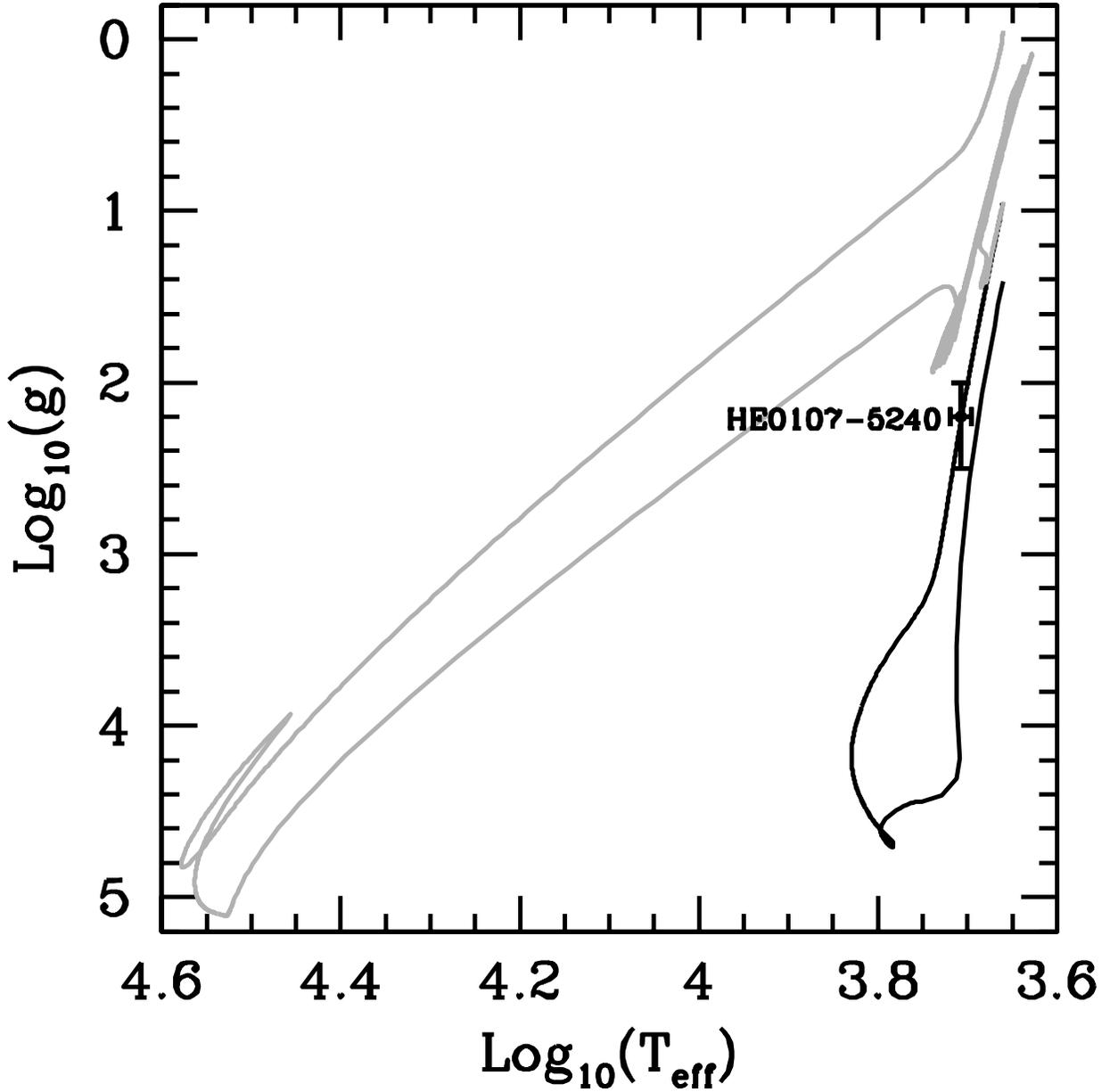}

\caption{Path followed by the 0.8 $\rm M_\odot$, Y=0.23 and $\rm Z=10^{\rm -7}$
from the contraction phase on the Hayashi track to the end of E-AGB phase. The gray
color mark the phases in which the surface chemical composition is enriched by the
products of the "hot" He-H mixing that occurs at the He core flash. The filled dot
represent the position of the most metal poor star presently known: HE0107-5240.
\label{f07}}

\end{figure}

The initial metallicity Z of this star at the moment of its formation
may be settled to $\rm Z=10^{\rm -7}$ if we assume (arbitrarily) a
scaled solar distribution of all the elements (and in particular of
the O) with respect to Fe. The key evolutionary features of a ${\rm
0.8~M_\odot}$ having this (scaled solar) metallicity are reported in
column 3 of Table 1 and show that this star lies within the range of
metallicities (actually $\rm Z_{\rm CNO}$) for which the "hot" He-H
mixing occurs. Therefore it is not necessary to invoke for this star
any, ad hoc, late accretion of polluted matter. In spite of this
encouraging first result, a comparison between the available data for
HE0107-5240 and the evolutionary properties of the stellar model
having $\rm Z=10^{\rm -7}$ shows that the mixing of H-rich matter in
the He convective shell during the He core flash cannot explain the
$\rm [C,N/Fe]>>0$ observed in this star. There are many reasons for
that, each of which would be sufficient by itself to exclude the
autopollution scenario. Let us analyze each of them separately.

a) Figure \ref{f07} shows the path followed by the
${\rm 0.8~M_\odot}$ having a scaled solar metallicity $\rm Z=10^{\rm -7}$
in the $\rm Log(g)-Log(T_{\rm eff})$ plane as a solid line. 
The black portion of the path refers to the phases in
which the surface chemical composition is still the pristine one, while
the gray path shows the part of track where the star shows the large
overabundances of the elements brought to the surface during the He
core flash. The position of HE0107-5240 is marked by the black
filled dot. It is strikingly evident that the position of HE0107-5240
is still on the middle of the first RGB, well below the He core flash
(and well below the central He burning location). Note that 
$\rm Log(g)=2$, as discussed by \citet{chris02}, is a firm upper
limit to the gravity of HE0107-5240, occurence that reinforces
such a finding.

b) Table 1 shows the time spent by the model on the first RGB (46 Myr
above $\rm Log(L/L_\odot)=2$), on the second RGB (1.4 Myr) and in the
central He burning phase (41 Myr, 14 Myr of which spent at effective
temperatures lower than 7000 K). On the basis of these lifetimes one
should expect to see at least 30 metal free Red Giants for each Red
Giant showing very large [C,N/Fe] ratios. None of them has been seen.
One could think, on the other hand, that HE0107-5240 is in the central
He burning phase, but also in this case we would expect roughly 3
metal free Red Giants for each auto polluted Red Giant. Once again
none of them has been discovered so far.

c) The surface [C/Fe] and [N/Fe] predicted by the models are,
respectively, 6.07 and 5.8, values that must be compared to the
observed ones: $\rm [C/Fe]_{\rm HE0107-5240}$=4 and $\rm [N/Fe]_{\rm
HE0107-5240}$=2.3. Also in this case the predicted values are
exceedingly large with respect to the observed ones. Even if we forget
about the Fe abundance and concentrate on the C/N ratio, we find that
the models predict a ratio roughly 1 while the observed ratio is
roughly 50.

d) The $\rm ^{\rm 12}C/^{\rm 13}C$ ratio predicted by the model is of
the order of 3:4 while the lower limit for the observed value is of
the order of 30.

e) The model predicts this star to be very Li rich, in particular
[Li/Fe]=5.5, while a (very) upper limit has been set for this star to
5.3.

Actually there is not even one point in favor of the auto pollution
scenario, so that we are forced to conclude that the surface chemical
composition of HE0107-5240 cannot be explained by the auto pollution
that occurs during the He core flash in the low mass extremely Iron
poor stars. In our opinion the "massive star" scenario presented by
\cite{blc03} and \cite{lcb03} still seems the most promising one to
explain not only this star but the bulk of the extremely Iron poor
stars: the ones showing strong Carbon and Nitrogen overabundances as
well as those not showing overabundances of these elements.

\acknowledgements

A.C. warmly thanks John Lattanzio and Brad Gibson for their kind hospitality in 
Melbourne and for having provided the computer facilities (the Australian 
Partnership for Advanced Computing National Facility and the Swinburne Centre for 
Astrophysics and Supercomputing in Melbourne) necessary to perform such a huge 
amount of computations.

\include{tab1}

\end{document}

%% file: tab1.tex
\begin{deluxetable}{lrrrrr}
\small
\tablenum{1}
\tablewidth{0pt}
\tablecaption{Surface abundances and key evolutionary properties} 
\tablehead{
\colhead{}        & \colhead{(1)}  & \colhead{(2)} & \colhead{(3)} & \colhead{(4)} & \colhead{(5)}
}
\startdata  
$\rm M(M_\odot)$      &      0.8                    &      0.8                    &      0.8                    &      0.8                    &      1.0                    \\
Z                     &      0                      & $\rm 1\times10^{\rm -8}$    & $\rm 1\times10^{\rm -7}$    &      0                      &      0                      \\   
Y                     &      0.23                   &      0.23                   &      0.23                   &      0.27                   &      0.23                   \\   
H                     &    0.4718                   &    0.4817                   &    0.4440                   &    0.3739                   &    0.5161                   \\
$\rm ^{\rm 3}He$      & $\rm 4.16\times10^{\rm -4}$ & $\rm 7.45\times10^{\rm -4}$ & $\rm 9.86\times10^{\rm -4}$ & $\rm 3.95\times10^{\rm -4}$ & $\rm 7.02\times10^{\rm -4}$ \\
$\rm ^{\rm 4}He$      &    0.5098                   &    0.4968                   &    0.5291                   &    0.6077                   &    0.4699                   \\
$\rm ^{\rm 7}Li$      & $\rm 2.45\times10^{\rm -6}$ & $\rm 5.44\times10^{\rm -7}$ & $\rm 7.63\times10^{\rm -7}$ & $\rm 1.95\times10^{\rm -7}$ & $\rm 3.90\times10^{\rm -8}$ \\
$\rm ^{\rm 12}C$      & $\rm 6.85\times10^{\rm -3}$ & $\rm 9.66\times10^{\rm -3}$ & $\rm 1.60\times10^{\rm -2}$ & $\rm 9.18\times10^{\rm -3}$ & $\rm 6.12\times10^{\rm -3}$ \\
$\rm ^{\rm 13}C$      & $\rm 1.97\times10^{\rm -3}$ & $\rm 2.47\times10^{\rm -3}$ & $\rm 3.47\times10^{\rm -3}$ & $\rm 2.15\times10^{\rm -3}$ & $\rm 1.90\times10^{\rm -3}$ \\
$\rm ^{\rm 14}N$      & $\rm 6.60\times10^{\rm -3}$ & $\rm 4.61\times10^{\rm -3}$ & $\rm 3.64\times10^{\rm -3}$ & $\rm 5.87\times10^{\rm -3}$ & $\rm 4.72\times10^{\rm -3}$ \\
$\rm ^{\rm 15}N$      & $\rm 7.47\times10^{\rm -6}$ & $\rm 2.10\times10^{\rm -7}$ & $\rm 1.47\times10^{\rm -7}$ & $\rm 1.36\times10^{\rm -6}$ & $\rm 1.75\times10^{\rm -7}$ \\
$\rm ^{\rm 16}O$      & $\rm 2.52\times10^{\rm -3}$ & $\rm 3.98\times10^{\rm -3}$ & $\rm 2.80\times10^{\rm -3}$ & $\rm 6.27\times10^{\rm -4}$ & $\rm 5.84\times10^{\rm -4}$ \\
$\rm ^{\rm 17}O$      & $\rm 1.35\times10^{\rm -5}$ & $\rm 1.53\times10^{\rm -5}$ & $\rm 9.37\times10^{\rm -6}$ & $\rm 2.59\times10^{\rm -6}$ & $\rm 2.57\times10^{\rm -6}$ \\
$\rm ^{\rm 18}O$      & $\rm 1.97\times10^{\rm -6}$ & $\rm 9.72\times10^{\rm -8}$ & $\rm 2.44\times10^{\rm -8}$ & $\rm 3.40\times10^{\rm -7}$ & $\rm 5.36\times10^{\rm -9}$ \\
$\rm ^{\rm 19}F$      & $\rm 3.44\times10^{\rm -8}$ & $\rm 1.08\times10^{\rm -8}$ & $\rm 6.31\times10^{\rm -9}$ & $\rm 6.37\times10^{\rm -9}$ & $\rm 3.10\times10^{\rm -9}$ \\
$\rm ^{\rm 20}Ne$     & $\rm 6.03\times10^{\rm -8}$ & $\rm 2.03\times10^{\rm -7}$ & $\rm 1.06\times10^{\rm -7}$ & $\rm 3.67\times10^{\rm -9}$ & $\rm 8.96\times10^{\rm -9}$ \\
$\rm ^{\rm 22}Ne$     & $\rm 6.06\times10^{\rm -9}$ & $\rm 3.74\times10^{\rm -8}$ & $\rm 4.22\times10^{\rm -8}$ & $\rm 4.30\times10^{\rm-10}$ & $\rm 6.98\times10^{\rm-10}$ \\
$\rm \Delta T~(1st~RGB)(Yr)$              & $\rm 38\times10^6$    & $\rm 47\times10^6$    & $\rm 46\times10^6$       & $\rm 32\times10^6$       & $\rm 33\times10^6$   \\
$\rm M_{\rm He}~(1st~Flash)(M_\odot)$     & 0.517                 & 0.543                 & 0.547                    & 0.496                    & 0.498                \\
$\rm M_{\rm Tmax}~(1st~Flash)(M_\odot)$   & 0.348                 & 0.375                 & 0.352                    & 0.256                    & 0.256                \\
$\rm Log(L/L_\odot)~(1st~Flash)$          & 2.45                  & 2.77                  & 2.98                     & 2.38                     & 2.37                 \\
$\rm \Delta T(Flash)~(1st~Flash)(Yr)$     & $\rm 2.24\times10^5$  & $\rm 1.3\times10^5$   & $\rm 1.5\times10^5$      & $\rm \simeq 1\times10^5$ & $\rm 6\times10^4$    \\
$\rm M_{\rm He}~(2nd~RGB)(M_\odot)$       & 0.357                 & 0.403                 & 0.362                    & $----$                   & 0.261                \\
$\rm M_{\rm Tmax}~(2nd~RGB)(M_\odot)$     & 0.347                 & 0.373                 & 0.350                    & $----$                   & 0.255                \\
$\rm Log(L/L_\odot)~(2nd~RGB)$            & 3.262                 & 3.477                 & 3.707                    & $----$                   & 3.051                \\
$\rm \Delta T_{\rm 2}~(2nd~RGB)(Yr)$      & $\rm 2.09\times10^6$  & $\rm 1.03\times10^6$  & $\rm 1.39\times10^6$     & $----$                   & $\rm 11.9\times10^6$ \\
$\rm M_{\rm He}~(2nd~Flash)(M_\odot)$     & 0.486                 & 0.490                 & 0.473                    & $----$                   & 0.463                \\
$\rm M_{\rm Tmax}~(2nd~Flash)(M_\odot)$   & 0.360                 & 0.357                 & 0.330                    & $----$                   & 0.232                \\
$\rm Log(L/L_\odot)~(2nd~Flash)$          & 3.726                 & 3.725                 & 3.693                    & $----$                   & 3.525                \\
$\rm M_{\rm He}~(beg.~He~burn.)(M_\odot)$ & 0.497                 & 0.499                 & 0.484                    & $----$                   & $----$               \\
$\rm M_{\rm He}~(end~He~burn.)(M_\odot)$  & 0.749                 & 0.749                 & 0.757                    & $----$                   & $----$               \\
$\rm \Delta T~(He~burn.)(Yr)$             & $\rm 43\times10^6$    & $\rm 43\times10^6$    & $\rm 41\times10^6$       & $----$                   & $----$               \\
\enddata             
\end{deluxetable}   